\documentstyle[11pt,newpasp,twoside]{article}
\begin{document}
\title{Determining the abundance of extragalactic planets}
\author{M. Dominik}
\affil{University of St Andrews,
School of Physics \& Astronomy,
North Haugh, St Andrews, KY16 9SS, United Kingdom}
\begin{abstract}
Gravitational microlensing provides a unique technique to reveal information
about extragalactic
planets. A network of at least four 2m-class telescopes distributed 
around the northern hemisphere could probe 15--35 jupiters and 4--10 saturns
around M31
stars per year.
\end{abstract}

\section*{Planet detection by microlensing}

By means of creating distortions to the gravitational field of their
parent stars, 
{\em unseen planets} of mass $m$ around {\em unseen lens stars} of
mass $M$ can cause
deviations of 1--20$\,$\% (lasting days for jupiters to hours for earths) 
to the light curves of background stars undergoing a microlensing event
which lasts $\sim\,$1~month.
The probability for a planetary signal varies
with the projected orbital radius of the planet $r_{\rm p}$ 
reaching a maximum near 
$r_{\rm p} \sim r_{\rm E}$, with 
the {\em Einstein radius} 
$r_{\rm E} = \sqrt{2\,R_{\rm S}\,D}$,
the Schwarzschild radius of the lens star $R_{\rm S} = (2GM)/c^2$, 
$D = D_{\rm L}\,(D_{\rm S}-D_{\rm L})/D_{\rm S}$,
and $D_{\rm S}$ and $D_{\rm L}$ being source or lens distances.

By achieving a galactic length scale 
$D \sim 2.5~\mbox{kpc}\;[r_{\rm E}/(2.5~\mbox{AU})]^2$
for M-dwarf lens stars ($M \sim 0.3~M_\odot$), one becomes sensitive
to planetary systems similar to our own. This condition is 
met not only for observations of Galactic bulge stars
being lensed by stars in both the Galactic disk or the bulge itself
($D_{\rm S} \sim 8.5~\mbox{kpc}$, $D_{\rm L} \sim 6~\mbox{kpc}$), 
but also for observations of stars in other galaxies as
M31 (Crotts 1992; Jetzer 1994) lensed by stars
in the same galaxy ($D_{\rm S} \gg D_{\rm S}-D_{\rm L}$, $D \approx D_{\rm L}$).

While the probability to detect a planetary deviation and its duration
decreases with smaller mass ratios $q = m/M$ 
(roughly as $\sqrt{q}$), the maximal 
deviation 
$\delta_{\rm th}$  
is only limited by the radius of the source star $R_\star$
(Bennett \& Rhie 1996).
For $\delta_{\rm th} = 10\,$\%
on observing M31 stars, 
a minimal mass ratio of
$q_{\rm min} = 2 \times 10^{-3}\,
[R_\star/(100~R_{\mbox{\sun}})]^2$
is required for $r_{\rm E} \sim 2.5~\mbox{AU}$ and M-dwarf lens stars,
corresponding to planets with twice the mass of Saturn for
$R_\star \sim 100~R_\odot$.

\section*{M31 microlensing observations}

Due to its small angular size, the monitoring of M31 requires only a few
fields and therefore only a fraction of telescope time during each night.
A proposed network of at 
least four 2m-class telescopes distributed around
the northern hemisphere (Dominik 2002) will provide a 
sampling interval of $\sim 6\,$h. 
{\em On-line data reduction} based on {\em difference-imaging}
will allow an increased sampling through 
target-of-oppurtunity observations in the case of ongoing anomalies.

Since M31 observations comprise 
{\em unresolved star fields}, only events with 
{\em high peak magnification} $A_0$ on {\em bright sources}
are detectable,
ruling out catching signatures of Earth-mass planets.
The smaller fraction of useful events is partly compensated 
by the {\em large number of source stars} and  
an {\em increased planet detection efficiency} for larger
peak magnifications (Griest \& Safizadeh 1998).   

The table compares the detection capabilities for
Galactic and extragalactic planets with ground-based campaigns.
Planet detection efficiencies follow Gould \& Loeb (1992) and
Covone et al.\ (2000) and are averaged over the lensing zone
($0.6~r_{\rm E} \leq r_{\rm p}
\leq 1.6~r_{\rm E}$).
The determination of 
planet parameters is not required for
obtaining upper abundance limits based on the absence
of deviations.

\begin{table}
\caption{Planet detection capabilities for an M31 network
and the PLANET campaign (Dominik et al.~2002)
on Galactic bulge stars}  
\footnotesize
\begin{tabular}{lcc}
\tableline
 & Galactic bulge  & M31 \\[0.3ex]
\tableline \\[-1.35ex]
number of source stars & $\sim\,10^{7}$ & $\sim\,10^{10}$\\[0.3ex]
resolution of source stars & resolved & unresolved \\[0.3ex]
telescope time & dedicated & 0.5--2.5~h per night \\[0.3ex]
field of view [sq deg] & 0.004--0.03 & 0.01--1 \\[0.3ex]
number of fields  &  & \\[-0.1ex] 
monitored during night  &  \raisebox{1.5ex}[-1.5ex]{$\sim 20$} & 
\raisebox{1.5ex}[-1.5ex]{1--8} \\[0.3ex] 
mean sampling interval & 1.5--2.5~h & 4--6~h \\[0.3ex]
total event rate $[\mbox{yr}^{-1}]$  & $\sim 300$--600& $\sim 150$--400 
\\[0.3ex]
useful types & giants & \\[-0.1ex]
of source stars 
& main-sequence stars & \raisebox{1.5ex}[-1.5ex]{giants} \\[0.3ex]
useful peak magnifications & 
$A_0 \ga 2$ & $A_0 \ga 10$ \\[0.3ex]
rate of useful events $[\mbox{yr}^{-1}]$ & $\sim 75$ & $\sim 35$--100 \\[0.3ex]
 & $\sim 20\,$\% (jupiters) & $\sim 35\,$\% (jupiters)  \\[-0.1ex]
\raisebox{1.5ex}[-1.5ex]{planet detection efficiency} &   $\sim 1.5\,$\% (earths)  &
 $\sim 10\,$\% (saturns) \\[0.3ex]
planet probing rate $[\mbox{yr}^{-1}]$ & 15--25 jupiters, 2--3 earths & 
15--35 jupiters, 4--10 saturns  \\[0.3ex]
upper limit on planetary  & 4--7$\,$\% (jupiters) & 
3--7$\,$\% (jupiters)\\[-0.1ex]
abundance within 3 years & $\sim\,40\,$\% (earths) & 
10--30$\,$\% (saturns)\\[0.3ex]
location of parent stars & Galactic disk and bulge & M31
 \\[0.3ex]
extraction of & & mostly difficult \\[-0.1ex] 
planet parameters &\raisebox{1.5ex}[-1.5ex]{fair in many cases}
 & or even impossible \\ 
\tableline
\tableline
\end{tabular}
\end{table}

\end{document}